\documentclass[runningheads]{llncs}
\usepackage[T1]{fontenc}
\usepackage{graphicx}
\usepackage{bm}
\usepackage{color}
\usepackage{amsmath}
\usepackage[misc,geometry]{ifsym} 

\begin{document}
\title{Boundary Difference Over Union Loss For Medical Image Segmentation}
\author{Fan Sun\inst{1} \and
Zhiming Luo\inst{1}\textsuperscript{(\Letter)} \and
Shaozi Li\inst{1}
}
%1{Sun, Fan}
%2{Luo, Zhiming}
%3{Li, Shaozi}

\authorrunning{Sun et al.}
\institute{Department of Artificial Intelligence, Xiamen University, Xiamen, Fujian, China\\
\textsuperscript{\Letter}Correspondences: \email{zhiming.luo@xmu.edu.cn}
%\url{http://www.springer.com/gp/computer-science/lncs}
}

% \author{Anonymous Submission: 1247}

\maketitle              % typeset the header of the contribution
\begin{abstract}
% Medical image segmentation plays a vital role in clinical diagnosis. However, most of the current losses for medical image segmentation mainly account for the overall segmentation results. On the other aspects, fewer losses have been proposed to guide boundary segmentation, or they need to be used in combination with different losses and have ineffective results. In this study, we develop a simple and effective loss, Boundary Difference over Union Loss (Boundary DoU Loss), to guide the segmentation of the boundary regions. Our Boundary DoU Loss only relies on region calculation, which is easy to implement, training stable, and does not have to be combined with any other losses. At the same time, we use the size  of the target to adaptively change the attention applied to the boundary. Experimental results on two segmentation datasets (i.e., ACDC and Synapse) with UNet, TransUNet, and Swin-UNet demonstrate the effectiveness of our proposed loss.
Medical image segmentation is crucial for clinical diagnosis. However, current losses for medical image segmentation mainly focus on overall segmentation results, with fewer losses proposed to guide boundary segmentation. Those that do exist often need to be used in combination with other losses and produce ineffective results. To address this issue, we have developed a simple and effective loss called the Boundary Difference over Union Loss (Boundary DoU Loss) to guide boundary region segmentation. It is obtained by calculating the ratio of the difference set of prediction and ground truth to the union of the difference set and the partial intersection set. Our loss only relies on region calculation, making it easy to implement and training stable without needing any additional losses. Additionally, we use the target size to adaptively adjust attention applied to the boundary regions. Experimental results using UNet, TransUNet, and Swin-UNet on two datasets (ACDC and Synapse) demonstrate the effectiveness of our proposed loss function. Code is available at https://github.com/sunfan-bvb/BoundaryDoULoss.
\end{abstract}
\section{Introduction}
Medical image segmentation is a vital branch of image segmentation ~\cite{long2015fully,chen2017rethinking,zhao2017pyramid,he2017mask,liu2018path,chen2018encoder}, and can be used clinically for segmenting human organs, tissues, and lesions. Deep learning-based methods have made great progress in medical image segmentation tasks and achieved good performance, including early CNN-based methods~\cite{ronneberger2015u,zhou2018unet++,isensee2018nnu,huang2020unet}, as well as more recent approaches utilizing Transformers~\cite{zhou2021nnformer,xu2021levit,hatamizadeh2022unetr,valanarasu2021medical,gao2021utnet}.

\begin{figure}[t]
\centering
\includegraphics[width=0.9\linewidth]{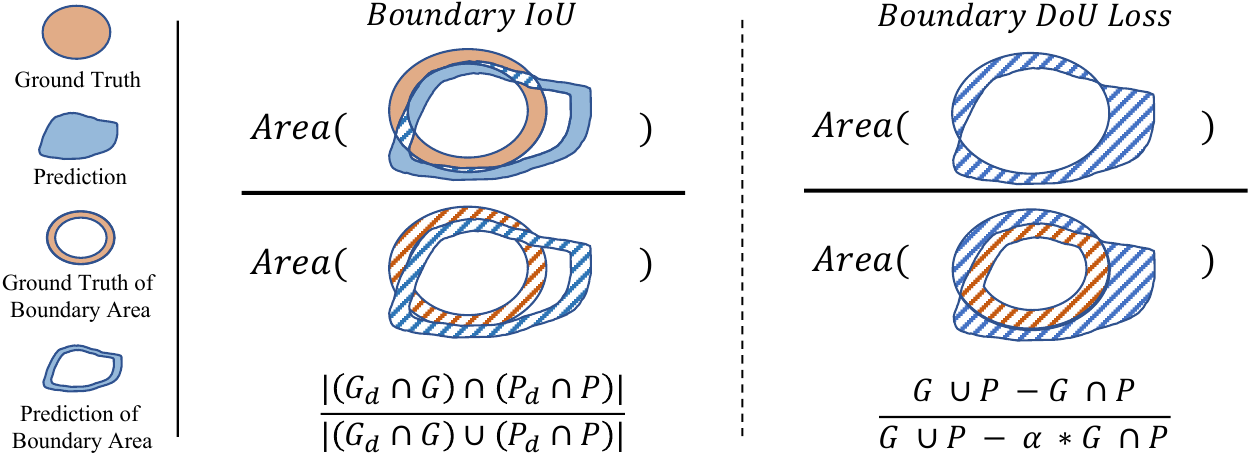}
\caption{The illustration of Boundary IoU (left) and Boundary DoU Loss (right), the shaded area in the figure will be calculated. $G$ and $P$ indicate ground truth and prediction, and $G_d$ and $P_d$ denote their correponding boundary areas. $\alpha$ is a hyper-parameter.}
\label{fig:b-iou}
\end{figure}

From CNN to Transformer, many different model architectures have been proposed, as well as a number of training loss functions. These losses can be mainly divided into three categories. The first class is represented by the Cross-Entropy Loss, which calculates the difference between the predicted probability distribution and the ground truth. Focal Loss~\cite{lin2017focal} is proposed for addressing hard-to-learn samples. The second category is Dice Loss and other improvements. Dice Loss~\cite{milletari2016v} is based on the intersection and union between the prediction and ground truth. Tversky Loss~\cite{salehi2017tversky} improves the Dice Loss by balancing precision and recall. The Generalized Dice Loss~\cite{sudre2017generalised} extends the Dice Loss to multi-category segmentation. The third category focuses on boundary segmentation. Hausdorff Distance Loss~\cite{karimi2019reducing} is proposed to optimize the Hausdorff distance, and the Boundary Loss~\cite{kervadec2019boundary} calculates the distance between each point in the prediction and the corresponding ground truth point on the contour as the weight to sum the predicted probability of each point. However, the current loss for optimizing segmented boundaries dependent on combining different losses or training instability. To address these issues, we propose a simple boundary loss inspired by the Boundary IoU metrics~\cite{cheng2021boundary}, \textit{i.e.}, Boundary DoU Loss.

Our proposed Boundary DoU Loss improves the focus on regions close to the boundary through a region-like calculation similar to Dice Loss. The error region near the boundary is obtained by calculating the difference set of ground truth and prediction, which is then reduced by decreasing its ratio to the union of the difference set and a partial intersection set. To evaluate the performance of our proposed Boundary DoU loss, we conduct experiments on the ACDC~\cite{bernard2018deep} and Synapse datasets by using the UNet~\cite{ronneberger2015u}, TransUNet~\cite{chen2021transunet} and Swin-UNet~\cite{cao2021swin} models. Experimental results show the superior performance of our loss when compared with others.

\section{Method}
\label{sec:pagestyle}

This section first revisit the Boundary IoU metric~\cite{cheng2021boundary}. Then, we describe the details of our Boundary DoU loss function and adaptive size strategy. Next, we discuss the connection between our Boundary DoU loss with the Dice loss. %Finally, we provide the final training loss function.

\subsection{Boundary IoU Metric}

The Boundary IoU is a segmentation evaluation metric which mainly focused on boundary quality. Given the ground truth binary mask $G$, the $G_d$ denotes the inner boundary region within the pixel width of $d$. The $P$ is the predicted binary mask, and $P_d$ denotes the corresponding inner boundary region, whose size is determined as a fixed fraction of 0.5\% relative to the diagonal length of the image. Then, we can compute the Boundary IoU metric by using following equation, as shown in the left of Fig.~\ref{fig:b-iou}, 
\begin{equation}
Boundary \ IoU = \frac{|(G_d\cap G)\cap(P_d\cap P)|}{|(G_d\cap G)\cup (P_d\cap P)|}.
\end{equation}
A large Boundary IoU value indicates that the $G_d$ and $P_d$ are perfectly matched, which means $G$ and $P$ are with a similar shape and their boundary are well aligned. In practice, the $G_d$ and $P_d$ is computed by the erode operation~\cite{cheng2021boundary}. However, the erode operation is non-differentiable, and we can not directly leverage the Boundary IoU as a loss function for training for increasing the consistency between two boundary areas.

% Boundary IoU is a boundary-specific evaluation metric that neither over-penalises nor ignores small targets. It focuses on the boundary region and calculates the intersection ratio of the two boundary regions of ground truth and prediction, which is shown in the left of Fig. 1. And the formula is
% \begin{equation}
% Boundary \ IoU = \frac{|(G_d\cap G)\cap(P_d\cap P)|}{|(G_d\cap G)\cup (P_d\cap P)|},
% \end{equation}
% where $G_d$ and $G_P$ denote the regions of ground truth and prediction respectively after taking a distance of d inwards. The part of $G_d$ that intersects $G$ is the edge of ground truth with width d, and the same for prediction. However, $G_d$ needs to be obtained by eroding $G$ and therefore cannot be backwards derived, so the Boundary IoU cannot be used directly as a loss, we therefore propose a method of approximating the two regions to calculate the loss.

\subsection{Boundary DoU Loss}
As shown in the left of Fig.~\ref{fig:b-iou}, we can find that the union of the two boundaries $|(G_d\cap G)\cup (P_d\cap P)|$ actually are highly correlated to the difference set between $G$ and $P$. The intersection $|(G_d\cap G)\cap(P_d\cap P)|$ is correlated to the inner boundary of the intersection of $G$ and $P$. If the difference set $G\cup P-G\cap P$ decreases and the $G\cap P$ increases, the corresponding Boundary IoU will increase.

Based on the above analysis, we design a Boundary DoU loss based on the difference region to facilitate computation and backpropagation. First, we directly treat the difference set as the miss-matched boundary between $G$ and $P$. Besides, we consider removing the middle part of the intersection area as the inner boundary, which is computed by $\alpha * G\cap P~(\alpha<1)$ for simplicity. Then, we joint compute the $G\cup P- \alpha *  G\cap P$ as the partial union. Finally, 
%We then incorporate the Boundary IoU idea of intersection over union for boundaries, and use all difference sets instead of difference sets for boundaries, which is more convenient for computation and can optimize all false positive and false negative regions at the same time.
as shown in the right of Fig.~\ref{fig:b-iou}, our Boundary DoU Loss can be computed by,
\begin{equation}
L_{DoU}= \frac{G\cup P-G\cap P}{G\cup P- \alpha *  G\cap P},
\end{equation}
where $\alpha$ is a hyper-parameter controlling the influence of the partial union area. 

%Boundary DoU refers to the ratio of the difference set of ground truth and prediction to the partial union set, which refers to the union of the difference set and the partial intersection set. Since only a small part of the intersection is taken, the proportion of the difference set is increased, and thus lowering the loss can make the difference set smaller faster.

\begin{figure}[t]
\centering
\includegraphics[width=0.7\linewidth]{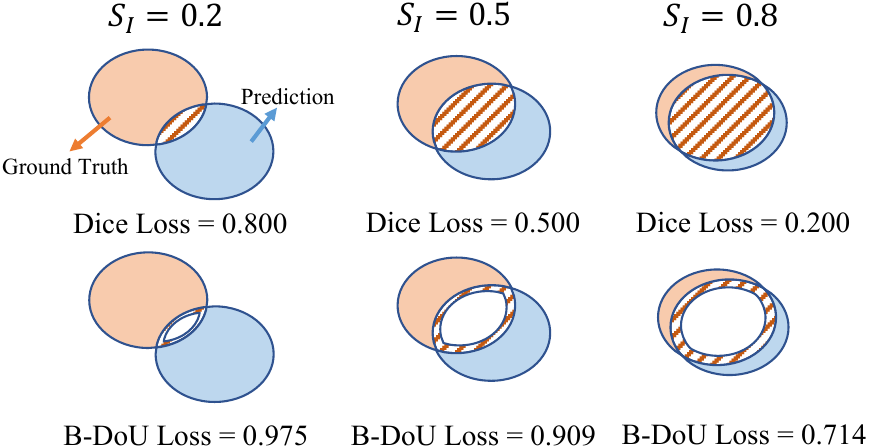}
\caption{Comparison of Boundary DoU Loss (B-DoU Loss) and Dice Loss. The figure shows the values of the two losses calculated at 20\%, 50\%, and 80\% of the intersection of Ground Truth and Prediction, respectively. We assume that both Ground Truth and Prediction have an area of 1 and $\alpha$ is 0.8.}
\label{fig:dice}
\end{figure}

\textbf{Adaptive adjusting $\alpha$ based-on target size:}
% The alpha is obtained by adapting from the target size and shape. Targets in medical images are often difficult to segment in the area close to the boundary. For large and simple shaped targets, the ratio of the edge area is small, while for small and complex shaped targets, the proportion is much larger. Therefore, we measure the size and shape of a target, i.e. its difficulty in segmentation, by the ratio of the edges to the complete target.
On the other aspect, the proportion of the boundary area relative to the whole target varies for different sizes. When the target is large, the boundary area only accounts for a small proportion, and the internal regions can be easily segmented, so we are encouraged to focus more on the boundary area. In such a case, using a large $\alpha$ is preferred. However, when the target is small, neither the interior nor the boundary areas are easily distinguishable, so we need to focus simultaneously on the interior and boundary, and a small $\alpha$ is preferred. To achieve this goal, we future adaptively compute $\alpha$ based on the proportion,
% For medical images, when the target is large, the internal regions can be easily segmented, so there is a greater need to focus on fine-tuning the area close to the boundary. And when the target is small, neither the interior nor the areas near the edges are easily distinguishable, so both should be attended to. 
\begin{equation}
	\alpha = 1 - 2\times\frac{C}{S}, \alpha\in [0, 1),
\end{equation}
where $C$ denotes the boundary length of the target, and $S$ denotes its size. 
% The ratio of the perimeter, which represents the number of pixels on the edge, to the area, which represents the number of pixels contained in the whole target, can measure the difficulty of segmenting the target.

 % In Eq. (2), loss is constant at 1 when $\alpha$ is 1, and degenerates to Dice Loss when $\alpha$ is -1. 
 % However, We argue that Dice Loss has too large a gradient for hard-to-segment samples and therefore causes it to be poorly optimised for hard-to-segment samples, because when the gradient is too large, it still changes by a large value when multiplied by the step size and therefore tends to fall into a local optimum. For the same step size, if the gradient is made lower, the model has more chances to explore better positions. Because larger targets have few hard-to-segment areas, they do not have a significant impact, while smaller targets have large hard-to-segment areas and therefore their performance on smaller targets is much worse. 
%  Therefore, we set $\alpha$ to be in $[-1, 1)$, which can strengthen the penalty on the area close to boundary, but also make the choice of different penalties for different targets. And since the value range of $\phi$ is $(0,1)$, we set the $\alpha$ to
% \begin{equation}
% 	\alpha = 2 * \phi - 1.
% \end{equation}

% Based on the statistic in the training dataset, we find that $C/S$ is within the range of [0, 0.5].

% Since the $C/S$ of the target are experimentally found to be distributed between 0 and 0.5, in order to widen the variation space we set $\alpha$ to
% \begin{equation}
% 	\alpha = 2 * \phi.
% \end{equation}

\subsection{Discussion}

%It may not be possible to tell from the formula whether it really works or not, and it is difficult to see at a glance whether it makes the segmentation better when the loss decreases. Therefore, we give the following proof.

In this part, we compare Boundary DoU Loss with Dice Loss. Firstly, we can re-write our Boundary DoU Loss as:
\begin{equation}
L_{DoU}= \frac{S_D}{S_D+S_I - \alpha S_I} = 1 - \frac{\alpha' * S_I}{S_D+\alpha'*S_I},
\end{equation}
where $S_D$ denotes the area of the difference set between ground truth and prediction, $S_I$ denotes the intersection area of them, and $\alpha'=1-\alpha$. Meanwhile, the Dice Loss can be expressed by the following: 
\begin{equation}
L_{Dice} = 1-\frac{2*TP}{2*TP+FP+FN}=1-\frac{2*S_I}{2*S_I+S_D},
\end{equation}
where TP, FP and FN denote True Positive, False Positive, and False Negative, respectively. 
It can be seen that Boundary DoU Loss and Dice loss differ only in the proportion of the intersection area. Dice is concerned with the whole intersection area, while Boundary DoU Loss is concerned with the boundary since $\alpha<1$. Similar to the Dice loss function, minimizing the $L_{DoU}$ will encourage an increase of the intersection area ($S_I\uparrow$) and a decrease of the different set ($S_D\downarrow$). Meanwhile, the $L_{DoU}$ will penalize more over the ratio of ${S_D}/{S_I}$.
%There are three scenarios when the loss drops. The first is when $S_D$ falls, at which point the difference set becomes smaller, and dice lifts at that moment. Second, when $S_D$ is constant, $S_I$ rises, at which the difference set is constant and the intersection set becomes larger, and dice rises. Third, when the $S_I$ rises faster than the $S_D$, the intersection set becomes larger than the difference set, the dice is raised. In all three cases, the segmentation result becomes better.
To corroborate its effectiveness more clearly, we compare the values of $L_{Dice}$ and $L_{DoU}$ in different cases in Fig.~\ref{fig:dice}. The $L_{Dice}$ decreases linearly with the difference set, whereas $L_{DoU}$ will decrease faster when $S_I$ is higher enough.

\section{Experiments}
\subsection{Datasets and Evaluation Metrics}

\textbf{Synapse:\footnote{\url{https://www.synapse.org/\#!Synapse:syn3193805/wiki/217789}}} The Synapse dataset contains 30 abdominal 3D CT scans from the  MICCAI 2015 Multi-Atlas Abdomen Labeling Challenge. Each CT volume contains $86\sim 198$ slices of $512\times512$ pixels. The slice thicknesses range from 2.5~mm to 5.0~mm, and in-plane resolutions vary from $0.54 \times 0.54~mm^2$ to $0.98 \times 0.98~mm^2$. Following the settings in TransUNet~\cite{chen2021transunet}, we randomly select 18 scans for training and the remaining 12 cases for testing.

\noindent\textbf{ACDC:\footnote{\url{https://www.creatis.insa-lyon.fr/Challenge/acdc/}}} The ACDC dataset is a 3D MRI dataset from the Automated Cardiac Diagnosis Challenge 2017 and contains cardiac data from 150 patients in five categories. Cine MR images were acquired under breath-holding conditions, with slices 5-8 mm thick, covering the LV from basal to apical, with spatial resolutions from 1.37 to 1.68 mm/pixel and 28 to 40 images fully or partially covering the cardiac cycle. Following the TransUNet, we split the original training set with 100 scans into the training, validation, and testing sets with a ratio of 7:1:2.

\noindent\textbf{Evaluation Metrics:} 
%We use the Dice Similarity Coefficient(DSC) and Hausdorff Distances(HD), which are most commonly used in medical image segmentation, as evaluation metrics. DSC focuses on segmentation results inside the target, while Hausdorff Distances focus on segmentation results of the boundaries. Besides, the Boundary IoU~\cite{cheng2021boundary}(Denoted as B-IoU) is adopted as another evaluation metric for the boundary, which calculates the intersection-over-union for mask pixels within a certain distance from the corresponding ground truth or prediction boundary contours. It is defined as
We use the most widely used Dice Similarity Coefficient (DSC) and Hausdorff Distances (HD) as evaluation metrics. Besides, the Boundary IoU~\cite{cheng2021boundary} (B-IoU) is adopted as another evaluation metric for the boundary.
% which calculates the intersection-over-union for pixels within a certain distance of the contours in the ground truth and the prediction. The B-IoU is defined as,
% \begin{equation}
% 	\mbox{B-IoU}(G, P) = \frac{|(G_d\cap G) \cap (P_d \cap P)|}{|(G_d\cap G) \cup (P_d \cap P)|},
% \end{equation}
% where $G$ is the ground truth, $P$ is the prediction, and $G_d$ and $P_d$ are the sets of all pixels within the distance of $d$ pixels from the contours of the ground truth and prediction, respectively.

% \begin{figure*}[t]
% \centering
% \includegraphics[width=0.95\linewidth]{compares.pdf}
% %  \includepdf[pages=-]{compare.pdf}
% \caption{The qualitative comparision of the segmentation results.
% (Row 1\&2: TransUNet on ACDC, Row 3\&4: Swin-UNet on ACDC, and Row 5\&6: TransUNet on CMR).
% %The first two rows show the results of the ACDC dataset on the TransUNet model, and the middle two rows indicate the results of the SwinUNet model for the ACDC dataset. The last two rows are the results for the CMR dataset. It is clear that our segmentation results, especially on the edges, outperform the other losses.
% }
% \label{fig:vis}
% \end{figure*}

\begin{table*}[t]
\centering
\caption{Experimental results on the Synapse dataset with the three models (average Dice score \% and average Hausdorff Distance in mm, and average Boundary IoU \%).
}
\label{tab:synapse}
\resizebox{\textwidth}{!}{
\begin{tabular}{l | c  c  c | c  c  c | c  c  c}
\hline %[2pt]设置线宽     
Model &  \multicolumn{3}{c|}{UNet} &  \multicolumn{3}{c|}{TransUNet} & \multicolumn{3}{c}{Swin-UNet} \\
\hline %[2pt]
Loss & DSC$\uparrow$ & HD$\downarrow$ & B-IoU$\uparrow$ & DSC$\uparrow$ & HD$\downarrow$ & B-IoU$\uparrow$ & DSC$\uparrow$ & HD$\downarrow$ & B-IoU$\uparrow$\\ %换行
\hline %[2pt]  
Dice & 76.38 & $31.45\pm 9.31$ & 86.26 & 78.52 & $28.84 \pm 2.47$ & 87.34 & 77.98 & $25.95\pm 9.07$ & 86.19  \\
 CE & 65.95 & $40.31\pm 50.3$ & 82.69 & 72.98 & $35.05 \pm 14.1$ & 84.84 & 71.77 & $33.20\pm 4.02$ & 84.04 \\
 Dice + CE & 76.77 & $30.20\pm 5.10$ & 86.21 & 78.19 & $29.30 \pm 11.8$ & 87.18 & 78.30 & $24.71\pm 2.84$ & 86.72 \\
 Tversky & 63.61 & $65.12 \pm 4.38$ & 75.65 & 63.90 & $70.89 \pm 218$ & 70.77 & 68.22 & $41.22\pm 419$ & 79.53 \\
 Boundary & 76.23 & $34.54\pm 12.5$ & 85.75 & 76.82 & $31.88 \pm 3.01$ & 86.66 & 76.00 & $26.74\pm 2.18$ & 84.98 \\
\hline %[2pt]  
 % Ours(dice based only) & 78.36 & {\bfseries23.66} & 87.26 & 78.28 & 30.91 & 87.12 & 79.37 & 24.07 & 87.34\\
 % Ours & {\bfseries79.76} & 27.25 & {\bfseries87.58} &{\bfseries79.26} &{\bfseries27.19} &  {\bfseries88.43} & {\bfseries79.87} & {\bfseries20.94} &  {\bfseries87.58} \\
 Ours & {\bfseries78.68} & \bm{$26.29\pm 2.35$} & {\bfseries87.08} & {\bfseries79.53} & \bm{$27.28\pm 0.51$} & {\bfseries88.11} & {\bfseries79.87} & \bm{$19.80\pm 2.34$} & {\bfseries87.78} \\
\hline %[2pt]     
\end{tabular}}
\end{table*}

\begin{table*}[t]
\centering
\caption{Experimental results on the ACDC dataset with the three models
(average Dice score \% and average Hausdorff Distance in mm, and average Boundary DoU \%).
}
\label{tab:acdc}
\resizebox{\textwidth}{!}{
\begin{tabular}{ l | c  c  c | c  c  c | c  c  c }
\hline %[2pt]设置线宽     
Model &  \multicolumn{3}{c|}{UNet} &  \multicolumn{3}{c|}{TransUNet} & \multicolumn{3}{c}{Swin-UNet} \\
\hline %[2pt]
Loss & DSC$\uparrow$ & HD$\downarrow$ & B-IoU$\uparrow$ & DSC$\uparrow$ & HD$\downarrow$ & B-IoU$\uparrow$ & DSC$\uparrow$ & HD$\downarrow$ & B-IoU$\uparrow$\\ %换行
\hline %[2pt]  
Dice & 90.17 & $1.34 \pm 0.10$ & 75.20 & 90.69 & $2.03 \pm 0.00$ & 76.66 & 90.17 & $1.34 \pm 0.01$ & 75.20  \\
 CE & 88.08 & $1.48 \pm 0.28$ & 71.25 & 89.22 & $2.07 \pm 0.05$ & 73.78 & 88.08 & $1.48 \pm 0.03$ & 71.25 \\
 Dice + CE & 89.94 & \bm{$1.28 \pm 1.29$} & 74.80 & 90.48 & $1.94 \pm 0.00$ & 76.27 & 89.94 & \bm{$1.28 \pm 0.00$} & 74.80 \\
 Tversky & 83.60 & $9.88 \pm 226$ & 69.36 & 90.37 & $1.95 \pm 0.02$ & 76.20 & 89.55 & $1.48 \pm 0.04$ & 74.37\\
 Boundary & 89.25  & $2.28 \pm 0.24$ & 73.08 & 90.48 &\bm{$1.84 \pm 0.08$} & 76.31 & 88.95 & $1.53 \pm 0.03$ & 72.73\\
\hline %[2pt]  
 % Ours(dice based only) & - & - & - & 90.65 & 1.95 & 76.36 & 89.89 & \textbf{1.30} & 75.08\\
 % Ours & {\bfseries91.05} & 2.46 & {\bfseries78.04} & {\bfseries91.27} & 1.84 & {\bfseries78.30} & {\bfseries90.31} & 1.56  & {\bfseries75.24}\\
Ours & {\bfseries90.84} & $1.54 \pm 0.33$ & {\bfseries76.44} & {\bfseries91.29} & $2.16 \pm 0.02$ & {\bfseries78.45} & {\bfseries91.02} & \bm{$1.28\pm 0.00$} & {\bfseries 77.00}\\

\hline %[2pt]     
\end{tabular}
}
\end{table*}

\begin{table*}[t]
\centering
\caption{Experimental results for different target sizes on the ACDC and Synapse datasets (average Dice score\%).
}
\label{tab:size}
\resizebox{\textwidth}{!}{
\begin{tabular}{ l | c  c | c  c | c  c | c  c | c  c | c  c }
\hline 
 &  \multicolumn{6}{c|}{ACDC} &  \multicolumn{6}{c}{Synapse} \\
\hline %[2pt]设置线宽     
Model &  \multicolumn{2}{c|}{UNet} &  \multicolumn{2}{c|}{TransUNet} & \multicolumn{2}{c|}{Swin-UNet} & \multicolumn{2}{c|}{UNet} &  \multicolumn{2}{c|}{TransUNet} & \multicolumn{2}{c}{Swin-UNet} \\
\hline %[2pt]
Loss & large & small & large & small & large & small &  large & small & large & small & large & small\\ %换行
\hline %[2pt]  
Dice & 92.60 & 84.11 & 93.63 & 85.95 & 92.40 & 86.43 & 78.59 & 36.22 & 80.84 & 36.16 & 79.97 & 41.67 \\
 CE & 91.40 & 81.72 & 92.55 & 83.86 & 91.49 & 82.58 & 68.89 & 0.12 & 75.21 & 32.32 & 74.27 & 26.25 \\
 Dice + CE & 92.91 & 84.36 & 93.45 & 85.70 & 92.83 & 85.30 & 78.89 & 38.27 & 80.47 & 36.82 & 80.29 & 42.22 \\
 Tversky & 91.82 & 70.37 & 93.28 & 85.86 & 92.67 & 84.53 & 65.76 & 24.60 & 66.00 & 25.74 & 70.08 & 34.44 \\
 Boundary & 92.53  & 83.97 & 93.53 & 85.57 & 92.25 & 83.63 & 78.37 &  37.30 & 79.00 & 37.14 & 78.02 & 39.23 \\
\hline %[2pt]  
 Ours & {\bfseries93.73} & {\bfseries86.04} & {\bfseries94.01} & {\bfseries86.93} & {\bfseries93.63} & {\bfseries86.83} & {\bfseries80.88} & {\bfseries38.68} & {\bfseries81.85} & {\bfseries37.32} & {\bfseries81.83} & {\bfseries44.08} \\
\hline %[2pt]     
\end{tabular}}
\end{table*}

\subsection{Implementation Details}
% For alpha, based on ablation experiments with $\alpha$ taking values from 0.1 to 1, we find that performance decreases significantly when the alpha takes values greater than 0.8, so we truncated alpha to 0.8 and the alpha value domain becomes [-1, 0.8].
% \begin{equation}
% 	\alpha = min(\alpha, 0.8)
% \end{equation}

%We conduct experiments on three advanced models to evaluate the performance of our proposed adaptive Dice loss, \textit{i.e.}, UNet, TransUNet and SwinUNet. The models are implemented with the PyTorch(version 1.10.2 with python 3.6) toolbox and run on an NVIDIA GTX A4000 GPU. The input resolution is set as $224\times 224$ for both datasets. For the SwinUNet and TransUNet networks, we used the same parameters as the original model, i.e. 150 epochs were trained on the Synapse dataset, and the ACDC dataset was validated on each epoch based on the 150 epochs trained. For the UNet, following the UNet setting in TransUNet, we choose resnet50 as the backbone and initialize the encoder with the ImageNet pre-trained weights. Otherwise, the configuration is the same as TransUNet. 
We conduct experiments on three advanced models to evaluate the performance of our proposed Boundary DoU Loss, \textit{i.e.}, UNet, TransUNet, and Swin-UNet. The models are implemented with the PyTorch toolbox and run on an NVIDIA GTX A4000 GPU. The input resolution is set as $224\times 224$ for both datasets. For the Swin-UNet~\cite{cao2021swin} and TransUNet~\cite{chen2021transunet}, we used the same training and testing parameters provided by the source code, \textit{i.e.}, the learning rate is set to 0.01, with a weight decay of 0.0001. The batch size is 24, and the optimizer uses SGD with a momentum of 0.9. For the UNet, we choose ResNet50 as the backbone and initialize the encoder with the ImageNet pre-trained weights following the setting in TransUNet~\cite{chen2021transunet}. The other configurations are the same as TransUNet. We train all models by 150 epochs on both Synapse and ACDC datasets.

%We also have detailed settings for the use of different comparison losses. Referring to the best results of the paper, we set $\alpha=0.7$ and $\beta=0.3$ in the Tversky Loss. For Boundary Loss, according to its paper and code, we made it combine with Dice Loss and Cross Entropy Loss, with a final $loss = \alpha * (Dice Loss + Cross Entropy Loss) + (1 - \alpha) * Boundary Loss$, where $\alpha$ is initially 1 and decreases by 0.01 each epoch until it equals 0.01 and then remains unchanged. In the experiments with Dice Loss + CE Loss, we retained the weight values of the original model, with TransUNet taken as (0.5, 0.5) and SwinUNet as (0.6, 0.4), while UNet is kept in the same configuration as TransUNet. Hausdorff Distance Loss, either alone or with Dice Loss, can't be optimized for both datasets and is therefore not included in the experimental results. 

We further train the three models by different loss functions for comparison, including Dice Loss, Cross-Entropy Loss (CE), Dice+CE, Tversky Loss~\cite{salehi2017tversky}, and Boundary Loss~\cite{kervadec2019boundary}. The training settings of different loss functions are as follows. For the $\lambda_1 \mbox{Dice}+\lambda_2 \mbox{CE}$, we set $(\lambda_1, \lambda_2)$ as (0.5, 0.5) for the UNet and TransUNet, and (0.6, 0.4) for Swin-UNet. For the Tversky Loss, we set $\alpha=0.7$ and $\beta=0.3$ by referring to the best performance in \cite{salehi2017tversky}. Following the Boundary Loss~\cite{kervadec2019boundary}, we use  $L = \alpha * (\mbox{Dice}+ \mbox{CE}) + (1 - \alpha) * \mbox{Boundary}$ for training . The $\alpha$ is initially set to 1 and decreases by 0.01 at each epoch until it equals 0.01.

% \begin{figure}[ht]
% \centering
% \includegraphics[width=0.7\linewidth]{test.pdf}
% \caption{The parameter sensitivity analysis of $\alpha$ on the ACDC dataset with TransUNet.
% \label{fig:alpha}
% }
% \end{figure}

\begin{figure}[t]
\includegraphics[width=\textwidth]{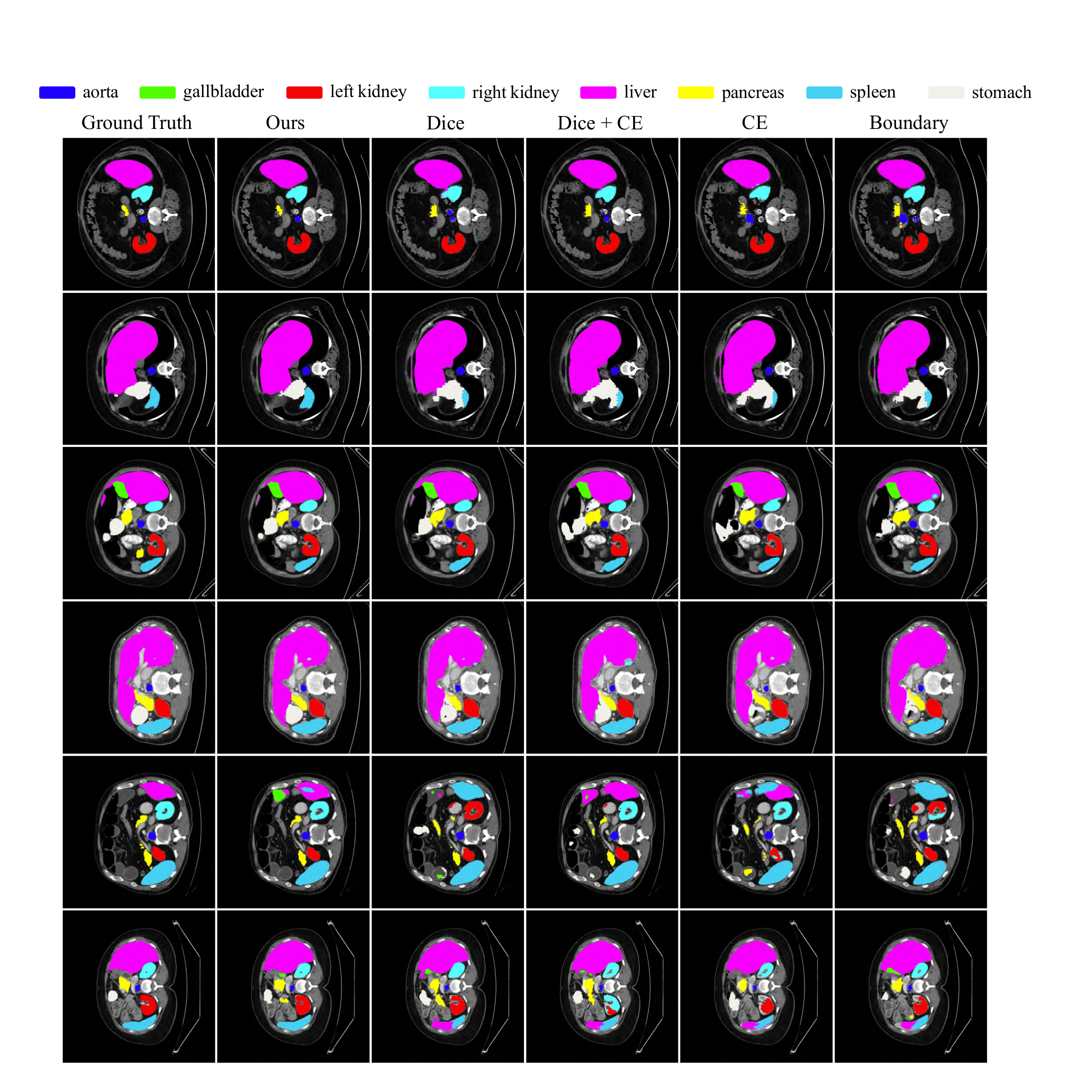}
\caption{The qualitative comparison of the segmentation on the Synapse dataset. (Row 1\&2: Swin-UNet, Row 3\&4: TransUNet, and Row 5\&6: UNet)} \label{synapse}
\end{figure}

\begin{figure}[t]
\includegraphics[width=\textwidth]{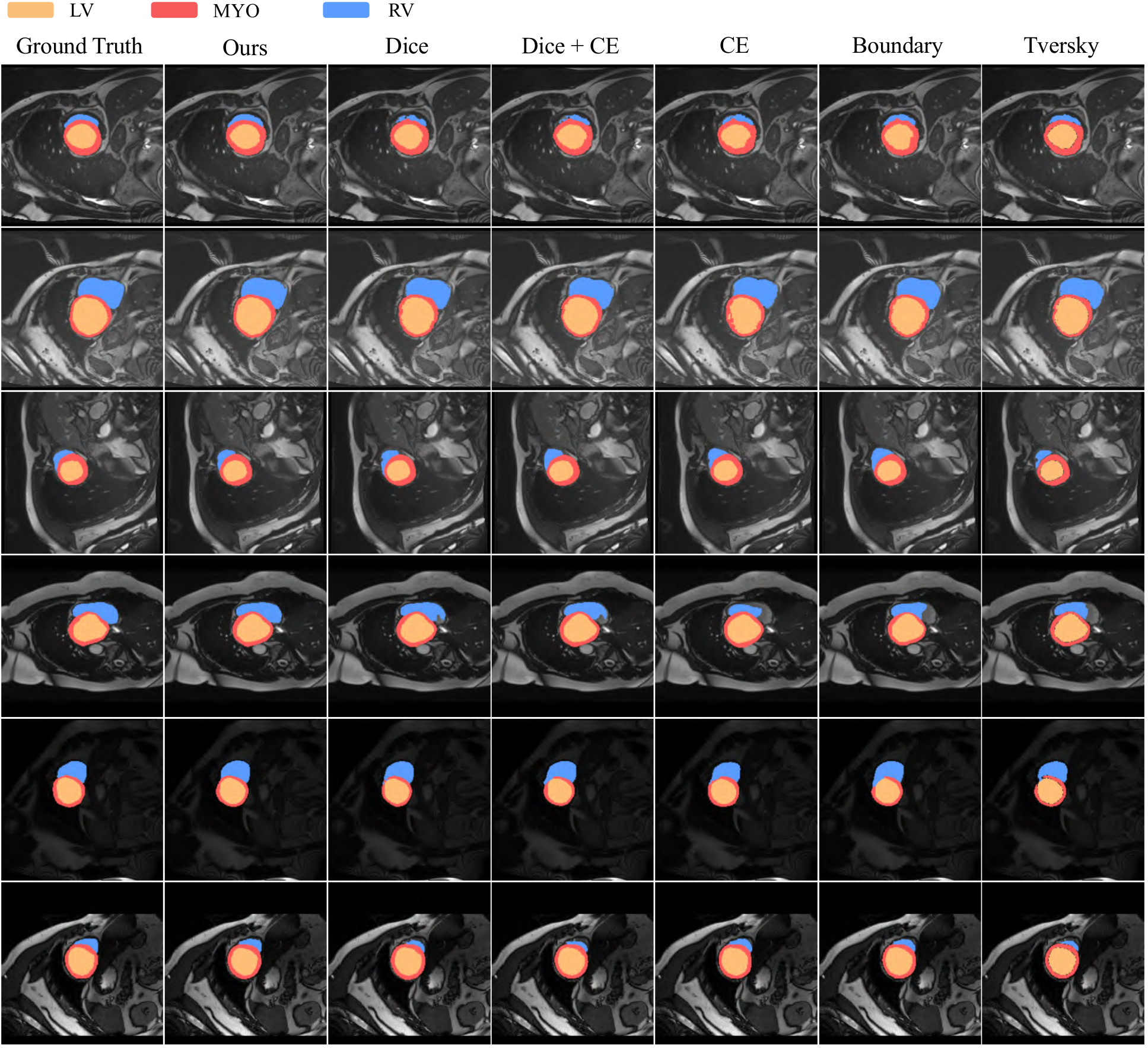}
\caption{The qualitative comparison of segmentation results on the ACDC dataset. (Row 1\&2: Swin-UNet, Row 3\&4: TransUNet, and Row 5\&6: UNet)} \label{acdc}
\end{figure}

\subsection{Results}
%In this part, we first conduct ablation experiments for alpha sampling values, and then report the performance of our proposed Boundary DoU Loss and other loss functions. All experimental results are obtained by the average value of five times running.

% {\color{blue}
% \textbf{Parameter Sensitivity Analysis:}
% In this section, we conduct a parameter sensitivity analysis of the $\alpha$ on the ACDC dataset using the TransUNet. The results of $\alpha$ in the range of [0.1, 0.9] are given in Fig.~\ref{fig:alpha}, and we can have following observations. \textbf{1)} The dice score is increased and Hausdorff Distance is decreased along with the increse of $\alpha$. The overall best performance is obtained when $\alpha=0.8$. \textbf{2)} When further increase the $\alpha$ to 0.9, the performance is significantly decreased.This is mainly due to a extremely larger $\alpha$ will results an over-representation of the boundary regions, and the attention to the inner regions will be decreased.
% }

\noindent\textbf{Quantitative Results:} Table~\ref{tab:synapse} shows the results of different losses on the Synapse dataset. From the table, we can have the following findings: 1) The original Dice Loss achieves the overall best performance among other losses. The CE loss function obtains a significantly lower performance than the Dice. Besides, the Dice+CE, Tversky, and Boundary do not perform better than Dice. 2) Compared with the Dice Loss, our Loss improves the DSC by 2.30\%, 1.20\%, and 1.89\% on UNet, TransUNet, and Swin-UNet models, respectively. The Hausdorff Distance also shows a significant decrease. Meanwhile, we achieved the best performance on the Boundary IoU, which verified that our loss could improve the segmentation performance of the boundary regions.

%Table~\ref{tab:acdc} reports the results for the ACDC dataset on the three models. Adaptive Loss improved the performance of DSC effectively on all three models. Compared with Dice Loss, UNet, TransUNet and SwinUNet improved by 0.62\%, 0.6\% and 0.85\% respectively. As the accuracy of the ACDC dataset is high enough, the small boost also demonstrates the capability of our loss. Since the Hausdorff Distance measures the maximum surface distance and does not reflect the average performance of the edges, although we don't get all optimal performance on it, we substantially outperformed all other losses on the Boundary IoU, indicating that our method is still better optimized in the regions close to the edges. Because we take a regional approach to optimization, there is no guarantee that the extremes are optimal, but they are certainly better at the mean. The results on both datasets also validate our reasoning that parts of medical images that are closer to the edges are more difficult to segment and that reducing the gradient allows for better optimization.

Table~\ref{tab:acdc} reports the results on the ACDC dataset. We can find that our Boundary DOU Loss effectively improves DSC on all three models. Compared with Dice Loss, the DSC of UNet, TransUNet, and Swin-UNet improved by 0.62\%, 0.6\%, and 0.85\%, respectively. Although our Loss did not get all optimal performance for the Hausdorff Distance, we substantially outperformed all other losses on the Boundary IoU. These results indicate that our method can better segment the boundary regions. This capability can assist doctors in better identifying challenging object boundaries in clinical settings.

\noindent\textbf{Qualitative Results:} Fig.~\ref{synapse} and ~\ref{acdc} show the qualitative visualization results of our loss and other losses. Overall, our method has a clear advantage for segmenting the boundary regions. In the Synapse dataset (Fig.~\ref{synapse}), we can achieve more accurate localization and segmentation for complicated organs such as the stomach and pancreas. Our results from the 3rd and 5th rows substantially outperform the other losses when the target is small. Based on the 2nd and last rows, we can obtain more stable segmentation on the hard-to-segment objects. As for the ACDC dataset (Fig.~\ref{acdc}), due to the large variation in the shape of the RV region as shown in Row 1, 3, 4 and 6, it is easy to cause under- or mis-segmentation. Our Loss resolves this problem better compared with other Losses. Whereas the MYO is annular and the finer regions are difficult to segment, as shown in the 2nd and 5th row, the other losses all result in different degrees of under-segmentation, while our loss ensures its completeness. Reducing the mis- and under-classification will allow for better clinical guidance.
 
% Fig.~\ref{synapse} shows the visualization results of the Synapse dataset on the three models. Lines two, three and four show that Adaptive loss is able to get more accurate segmentation in regions close to the edges, while the others show that it performs better on hard-to-segment samples.

% Fig.~\ref{acdc} represents the results of Adaptive Loss on the ACDC dataset compared to other losses. Since UNet has no results on a number of losses, we only show some of the losses for which TransUNet and SwinUNet are available. Obviously, we get better results with our loss on both large and small targets, especially in the areas near the boundary.

\noindent\textbf{Results of Target with Different Sizes:} 
We further evaluate the influence of the proposed loss function for segmenting targets with different sizes. Based on the observation of $C/S$ values for different targets, we consider a target to be a large one when $C/S < 0.2$ and otherwise as a small target. As shown in Table~\ref{tab:size}, our Boundary DoU Loss function can improve the performance for both large and small targets.

\section{Conclusion}
\label{sec:conlusion}
In this study, we propose a simple and effective loss (Boundary DoU) for medical image segmentation. It adaptively adjusts the penalty to regions close to the boundary based on the size of the different targets, thus allowing for better optimization of the targets. Experimental results on ACDC and Synapse datasets validate the effectiveness of our proposed loss function. 

\section*{Acknowledgement}
This work is supported by the National Natural Science Foundation of China (No. 62276221), the Natural Science Foundation of Fujian Province of China (No. 2022J01002), and the Science and Technology Plan Project of Xiamen (No. 3502Z20221025).

\bibliographystyle{splncs04}
\bibliography{mybibliography}

\begin{thebibliography}{10}
\providecommand{\url}[1]{\texttt{#1}}
\providecommand{\urlprefix}{URL }
\providecommand{\doi}[1]{https://doi.org/#1}

\bibitem{bernard2018deep}
Bernard, O., Lalande, A., Zotti, C., Cervenansky, F., Yang, X., Heng, P.A.,
  Cetin, I., Lekadir, K., Camara, O., Ballester, M.A.G., et~al.: Deep learning
  techniques for automatic mri cardiac multi-structures segmentation and
  diagnosis: is the problem solved? IEEE Transactions on Medical Imaging
  \textbf{37}(11),  2514--2525 (2018)

\bibitem{cao2021swin}
Cao, H., Wang, Y., Chen, J., Jiang, D., Zhang, X., Tian, Q., Wang, M.:
  Swin-unet: Unet-like pure transformer for medical image segmentation. arXiv
  preprint arXiv:2105.05537  (2021)

\bibitem{chen2021transunet}
Chen, J., Lu, Y., Yu, Q., Luo, X., Adeli, E., Wang, Y., Lu, L., Yuille, A.L.,
  Zhou, Y.: Transunet: Transformers make strong encoders for medical image
  segmentation. arXiv preprint arXiv:2102.04306  (2021)

\bibitem{chen2017rethinking}
Chen, L.C., Papandreou, G., Schroff, F., Adam, H.: Rethinking atrous
  convolution for semantic image segmentation. arXiv preprint arXiv:1706.05587
  (2017)

\bibitem{chen2018encoder}
Chen, L.C., Zhu, Y., Papandreou, G., Schroff, F., Adam, H.: Encoder-decoder
  with atrous separable convolution for semantic image segmentation. In:
  European Conference on Computer Vision (ECCV). pp. 801--818 (2018)

\bibitem{cheng2021boundary}
Cheng, B., Girshick, R., Doll{\'a}r, P., Berg, A.C., Kirillov, A.: Boundary
  iou: Improving object-centric image segmentation evaluation. In: IEEE/CVF
  Conference on Computer Vision and Pattern Recognition. pp. 15334--15342
  (2021)

\bibitem{gao2021utnet}
Gao, Y., Zhou, M., Metaxas, D.N.: Utnet: a hybrid transformer architecture for
  medical image segmentation. In: International Conference on Medical Image
  Computing and Computer-Assisted Intervention. pp. 61--71. Springer (2021)

\bibitem{hatamizadeh2022unetr}
Hatamizadeh, A., Tang, Y., Nath, V., Yang, D., Myronenko, A., Landman, B.,
  Roth, H.R., Xu, D.: Unetr: Transformers for 3d medical image segmentation.
  In: IEEE/CVF Winter Conference on Applications of Computer Vision. pp.
  574--584 (2022)

\bibitem{he2017mask}
He, K., Gkioxari, G., Doll{\'a}r, P., Girshick, R.: Mask r-cnn. In: IEEE
  International Conference on Computer Vision. pp. 2961--2969 (2017)

\bibitem{huang2020unet}
Huang, H., Lin, L., Tong, R., Hu, H., Zhang, Q., Iwamoto, Y., Han, X., Chen,
  Y.W., Wu, J.: Unet 3+: A full-scale connected unet for medical image
  segmentation. In: IEEE International Conference on Acoustics, Speech and
  Signal Processing (ICASSP). pp. 1055--1059. IEEE (2020)

\bibitem{isensee2018nnu}
Isensee, F., Petersen, J., Klein, A., Zimmerer, D., Jaeger, P.F., Kohl, S.,
  Wasserthal, J., Koehler, G., Norajitra, T., Wirkert, S., et~al.: nnu-net:
  Self-adapting framework for u-net-based medical image segmentation. arXiv
  preprint arXiv:1809.10486  (2018)

\bibitem{karimi2019reducing}
Karimi, D., Salcudean, S.E.: Reducing the hausdorff distance in medical image
  segmentation with convolutional neural networks. IEEE Transactions on Medical
  Imaging  \textbf{39}(2),  499--513 (2019)

\bibitem{kervadec2019boundary}
Kervadec, H., Bouchtiba, J., Desrosiers, C., Granger, E., Dolz, J., Ayed, I.B.:
  Boundary loss for highly unbalanced segmentation. In: International
  Conference on Medical Imaging with Deep Learning. pp. 285--296 (2019)

\bibitem{lin2017focal}
Lin, T.Y., Goyal, P., Girshick, R., He, K., Doll{\'a}r, P.: Focal loss for
  dense object detection. In: IEEE International Conference on Computer Vision.
  pp. 2980--2988 (2017)

\bibitem{liu2018path}
Liu, S., Qi, L., Qin, H., Shi, J., Jia, J.: Path aggregation network for
  instance segmentation. In: IEEE conference on Computer Vision and Pattern
  Recognition. pp. 8759--8768 (2018)

\bibitem{long2015fully}
Long, J., Shelhamer, E., Darrell, T.: Fully convolutional networks for semantic
  segmentation. In: IEEE Conference on Computer Vision and Pattern Recognition.
  pp. 3431--3440 (2015)

\bibitem{milletari2016v}
Milletari, F., Navab, N., Ahmadi, S.A.: V-net: Fully convolutional neural
  networks for volumetric medical image segmentation. In: International
  Conference on 3D Vision (3DV). pp. 565--571 (2016)

\bibitem{ronneberger2015u}
Ronneberger, O., Fischer, P., Brox, T.: U-net: Convolutional networks for
  biomedical image segmentation. In: International Conference on Medical Image
  Computing and Computer-Assisted Intervention. pp. 234--241 (2015)

\bibitem{salehi2017tversky}
Salehi, S.S.M., Erdogmus, D., Gholipour, A.: Tversky loss function for image
  segmentation using 3d fully convolutional deep networks. In: International
  Workshop on Machine Learning in Medical Imaging. pp. 379--387 (2017)

\bibitem{sudre2017generalised}
Sudre, C.H., Li, W., Vercauteren, T., Ourselin, S., Jorge~Cardoso, M.:
  Generalised dice overlap as a deep learning loss function for highly
  unbalanced segmentations. In: Deep learning in medical image analysis and
  multimodal learning for clinical decision support, pp. 240--248. Springer
  (2017)

\bibitem{valanarasu2021medical}
Valanarasu, J.M.J., Oza, P., Hacihaliloglu, I., Patel, V.M.: Medical
  transformer: Gated axial-attention for medical image segmentation. In:
  International Conference on Medical Image Computing and Computer-Assisted
  Intervention. pp. 36--46. Springer (2021)

\bibitem{xu2021levit}
Xu, G., Wu, X., Zhang, X., He, X.: Levit-unet: Make faster encoders with
  transformer for medical image segmentation. arXiv preprint arXiv:2107.08623
  (2021)

\bibitem{zhao2017pyramid}
Zhao, H., Shi, J., Qi, X., Wang, X., Jia, J.: Pyramid scene parsing network.
  In: IEEE conference on computer vision and pattern recognition. pp.
  2881--2890 (2017)

\bibitem{zhou2021nnformer}
Zhou, H.Y., Guo, J., Zhang, Y., Yu, L., Wang, L., Yu, Y.: nnformer: Interleaved
  transformer for volumetric segmentation. arXiv preprint arXiv:2109.03201
  (2021)

\bibitem{zhou2018unet++}
Zhou, Z., Rahman~Siddiquee, M.M., Tajbakhsh, N., Liang, J.: Unet++: A nested
  u-net architecture for medical image segmentation. In: Deep learning in
  medical image analysis and multimodal learning for clinical decision support,
  pp. 3--11. Springer (2018)

\end{thebibliography}

\end{document}